\documentclass[10pt]{article}
\usepackage{amsmath}
\usepackage{amssymb}
\usepackage{graphicx}

\def\aa{A\&A}

\def\aj{AJ}

\begin{document}

\setcounter{figure}{0}
\setcounter{table}{0}
\setcounter{footnote}{0}
\setcounter{equation}{0}

\vspace*{0.5cm}

\noindent {\Large \strut INTERACTION BETWEEN CELESTIAL AND TERRESTRIAL\\ \strut REFERENCE FRAMES AND SOME CONSIDERATIONS FOR\\
           \strut THE NEXT VLBI-BASED ICRF}
\vspace*{0.7cm}

\noindent\hspace*{1.5cm} Z. MALKIN$^1$, H. SCHUH$^2$, C. MA$^3$, S. LAMBERT$^4$\\
\noindent\hspace*{1.5cm} $^1$Pulkovo Observatory, St. Petersburg 196140, Russia\\
\noindent\hspace*{1.5cm} e-mail: malkin@gao.spb.ru\\
\noindent\hspace*{1.5cm} $^2$Institute of Geodesy and Geophysics, TU Vienna, 1040 Vienna, Austria\\
\noindent\hspace*{1.5cm} e-mail: harald.schuh@tuwien.ac.at\\
\noindent\hspace*{1.5cm} $^3$NASA Goddard Space Flight Center, Greenbelt MD 20771-0001, USA\\
\noindent\hspace*{1.5cm} e-mail: chopo.ma@nasa.gov\\
\noindent\hspace*{1.5cm} $^4$Paris Observatory, SYRTE, CNRS/UMR8630, 75014 Paris, France\\
\noindent\hspace*{1.5cm} e-mail: sebastien.lambert@obspm.fr\\

\vspace*{0.5cm}

\noindent {\large ABSTRACT.}
In this paper we outline several problems related to the realization of the international celestial and terrestrial reference frames ICRF and ITRF
at the millimeter level of accuracy, with emphasis on ICRF issues.
The main topics considered are: analysis of the current status of the ICRF, mutual impact of ICRF and ITRF, and some considerations for future ICRF realizations.

\vspace*{1cm}

\noindent {\large 1. INTRODUCTION}

\smallskip

International terrestrial and celestial reference frames, ITRF and ICRF, respectively, as well as the tie between them expressed by the Earth
Orientation parameters (EOP) are key products of geodesy and astrometry.
The requirements to all the components of this triad grows steadily and a mm and/or $\mu$as level of accuracy is the current goal of the astronomic and geodetic community.

The current computation procedures for ITRF and ICRF are based on multi-stage processing of observations made with several space geodetic techniques:
VLBI, SLR, GNSS, and DORIS.
Not all of them provide equal contributions to the final products.
The latest ITRF realizations have been derived from combination of normal equations obtained from all four techniques, whereas the ICRF is a result of a single global VLBI solution.
The latter is tied to the ITRF using an arbitrary set of reference stations.
But VLBI relies on the ITRF origin provided by satellite techniques and shares responsibility with SLR for the ITRF scale.
And all the techniques contribute to positions and velocities of ITRF stations.

This situation causes complicated mutual impact of ITRF and ICRF, which should be carefully investigated in order to improve the accuracy of both
reference systems and the consistency between each other and EOP.
The subject becomes more and more complicated when moving to millimeter accuracy in all components of this fundamental triad.
As a result, we have systematic errors in reference frames and inconsistency between them.

Due to its limited extent, this paper will deal only with the VLBI-derived ICRF errors and its dependence on the interaction with the ITRF.

\vspace*{0.7cm}

\noindent {\large 2. CURRENT ICRF STATUS}

\smallskip

\begin{figure}[t]
\centering
\includegraphics[clip,width=0.85\textwidth]{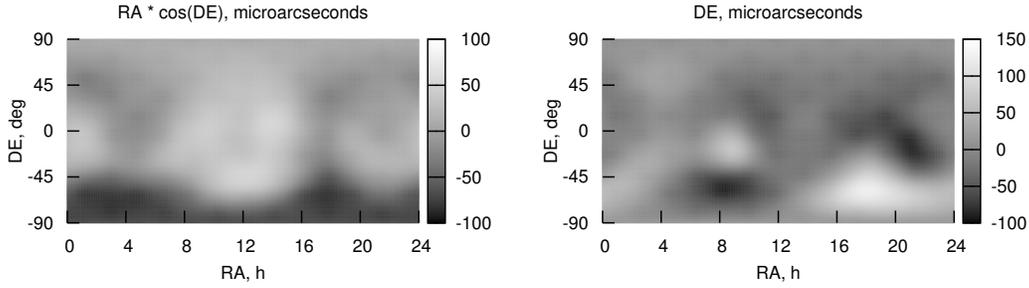}
\caption{Smoothed differences ICRF2 - ICRF in $\mu$as. One can see large (up to 200-250~$\mu$as depending on the smoothing) peak-to-peak differences normally attributed to the ICRF systematic errors.}
\label{fig:icrf2-icrf}
\end{figure}

The latest ICRF realization, ICRF2, was created in 2009 (Ma et al. 2009).
It provides much improvement as compared to the first ICRF (Ma et al. 1998, Fey et al. 2004) by:
\begin{itemize}
\itemsep = -0.7ex
\item increasing the total number of sources from 717 to 3414;
\item increasing the number of defining sources from 212 to 295;
\item improving the distribution of the defining sources over the sky;
\item improving the source position uncertainties (error floor) from 250~$\mu$as to 40~$\mu$as;
\item elimination of large systematic error at the level of about 0.2~mas (see Fig.\ref{fig:icrf2-icrf});
\item improving the axis stability from about 30~$\mu$as to about 10~$\mu$as.
\end{itemize}

\begin{figure}[t]
\centering
\includegraphics[clip,width=0.85\textwidth]{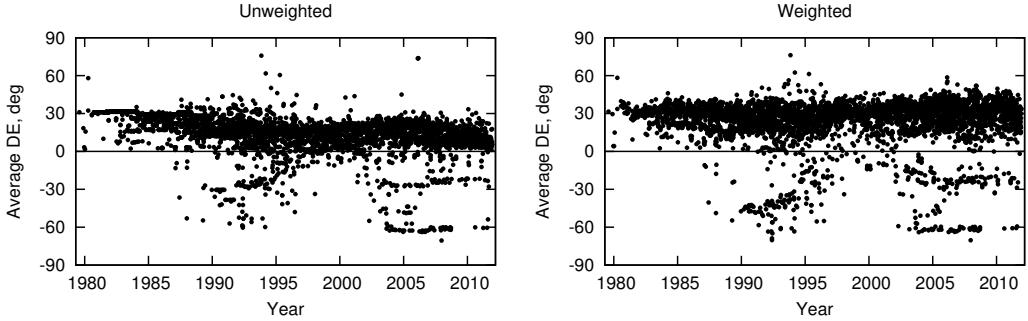}
\caption{Average declination obtained from 24h sessions: unweighted (left, simple averaging source declination for all observations)
and weighted (right, averaging declination with weights equal to the number of observations). One can see that the situation
does not much improve during the history of VLBI.}
\label{fig:de_mean}
\end{figure}

However, there are severe problems preventing further ICRF improvement, especially with respect to systematic errors.
Here is a list of some of them to be solved or mitigated in the next ICRF realizations:
\begin{itemize}
\itemsep = -0.7ex
\item the ICRS definition may need refinement;
\item the distribution of ICRF sources over the sky is rather uneven (weight on Northern sources, which is caused by the large number of Northern stations (see Fig.\ref{fig:de_mean});
\item the distribution of the errors in ICRF source positions over the sky is rather uneven;
\item proper (physical, coming from variable source structure) and apparent (coming from instrumental and analysis errors) source motions are
complicated and poorly predictable;
\item source positions depend on the wavelength (analogous to the color equation in optical astrometry);
\item computed source positions depend on the analysis options and observing network.
\end{itemize}

\vspace*{0.7cm}

\noindent {\large 3. MUTUAL IMPACT OF ICRF AND ITRF}

\smallskip

CRF results obtained from global VLBI solutions depend on the tie to the ITRF.
There are several problems that affect the source position catalog:
\begin{itemize}
\itemsep = -0.7ex
\item Dependence on the ITRF datum.
\item Dependence on the set of reference stations used. 
\item Dependence on the modeling of non-linear station motion.
\end{itemize}

The first problem seems to be not significant at the cm-level, but needs further investigation at the mm-level.
Selection of the set of reference (minimally constrained) stations that is used to tie VLBI global solutions to ITRF varies between different analysis centers and between different solutions obtained at the same center.

\begin{figure}[t]
\centering
\includegraphics[clip,width=0.85\textwidth]{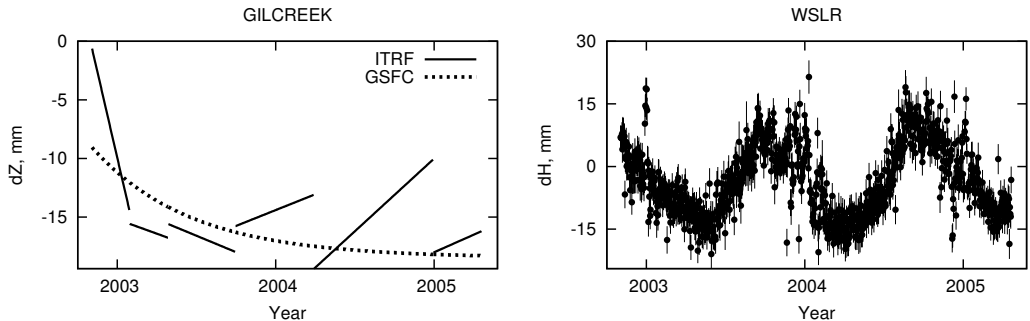}
\caption{Two examples of non-linear station movement. The plot on the left shows post-seismic displacements of VLBI station GILCREEK, AL, USA,
computed using the ITRF2008 model (solid line) and exponential model by MacMillan \& Cohen (2004).
The plot on the right shows seasonal height variations for IGS station WSLR, BC Canada as an example of large annual displacements.}
\label{fig:stations}
\end{figure}

Non-linear station movement represents a more serious problem that cannot be modeled properly in the ITRF.
ITRF describes station displacements only using a linear model with eventual jumps.
However, many stations show significant non-linear terms in position time series.
The most common are exponential movement of stations due to post-seismic relaxation and seasonal signals.
Two examples of such behavior are given in Fig.~\ref{fig:stations}.
For the WSLR station position, computed at SOPAC\footnote{http://sopac.ucsd.edu/}, one can see that using a linear IERS model is by no means satisfactory because this can lead to errors in station position modeling exceeding 1 cm.

This has two consequences. First, using these stations may disturb the ICRF orientation.
Secondly, if the actual station position for a given epoch differs substantially from the ITRF model, this increases both systematic errors and scatter of the EOP solution, which is obtained for this epoch from the daily set of observations involving the station concerned.
Both effects seem to be not carefully investigated yet.

Our goal is modeling station daily-averaged positions at mm-accuracy for any arbitrary epoch inside the operational
period, and for extrapolation to the real time and near future.
Several options are used in practice for this purpose:
\begin{enumerate}
\itemsep = -0.0ex
\item ITRF model: linear drift + jumps \\
  \indent --- not generally suitable
\item SOPAC model (Nikolaidis 2002): linear drift + jumps + seasonal term + exponential relaxation \\
  \indent --- performs much better \\
  \indent --- model parameters must be provided as a part of ITRF
\item GSFC model (Petrov 2005): B-splines \\
  \indent --- the best approximation to the computed movement \\
  \indent --- physical meaning is questionable \\
  \indent --- problems with dissemination, reproduction and extrapolation
\end{enumerate}

The second option is perhaps the most reasonable compromise between accuracy, predictability, and complexity.

\vspace*{0.7cm}

\noindent {\large 4. CONSIDERATIONS FOR FUTURE VLBI ICRF REALIZATIONS}

\smallskip

If one takes a look at the ICRF history, one can mention that the successive versions were issued with interval of about 5 years, see Table~\ref{tab:icrf_versions}. 
Following this sequence, we can set a goal to complete the next ICRF versions in 2014, and 2019.
It should be mentioned that the ICRF, ICRF-Ext.1, ICRF-Ext.2 represent, in fact, the same system based on unchanged coordinates of 212 defining sources.
Such long-time preserving of the positions of defining (and, as a rule, most observed) sources computed in 1995 may be the main reason of the ICRF
systematic errors discussed above (Sokolova \& Malkin 2007).
On the contrary, all the ICRF2 source positions were adjusted in a single global solution independently, and are tied to the ICRF only by orientation (NNR constraint).
One out of several options would be to keep this strategy for the next realizations too, and name them ICRF3 and ICRF4.
In this connection, perhaps it would be better to use the term 'core' sources instead of 'defining'. Other options for the calculation of the next ICRF realizations exist and will be examined thoroughly.

\begin{table}[t]
\centering
\caption{Realized (1995--2009) and prospective (2014, 2019) ICRF versions. For the latter, foreseen values are given in parentheses as a conservative guess.}
\medskip
\label{tab:icrf_versions}
\tabcolsep = 3.8pt
\begin{tabular}{lcccccccccccc}
ICRF version          && ICRF &   & ICRF-Ext.1 &   & ICRF-Ext.2 &   & ICRF2 &     & (ICRF3)  &     & (ICRF4) \\
\hline
Year                  && 1995 &   & 1999       &   & 2004       &   & 2009  &     & (2014)   &     & (2019) \\
Observations, mln     && 1.6  &   & 2.2        &   & 3.4        &   & 6.5   &     & (9--9.5) &     & (12--13) \\
\hline
Epoch difference, yr  &&      & 4 &            & 5 &            & 5 &       & (5) &          & (5) & \\
\end{tabular}
\end{table}

The primary goals of ICRF3 would be to incorporate new observations and enrich the set of core sources in the southern hemisphere.
It will be based on about 50\% more observations, and can help us to reach the following goals:
\begin{itemize}
\itemsep = -0.7ex
\item increase the total number of sources to $>$ 4100 (one source per 10 sq. deg.) mostly by southern sources;
\item increase number of the core sources to $>$ 410 (one source per 100 sq. deg.);
\item achieve a more uniform sky distribution of all and in particular core sources and position errors;
\item improve the source position errors, especially, in the Southern hemisphere. 
\end{itemize}

It is advisable to develop, simultaneously with the preparation of the ICRF3, new procedures of observation planning and an advanced strategy of station usage,
which shall allow us to reach the following goals with ICRF4 in 2019:
\begin{itemize}
\itemsep = -0.7ex
\item substantially increase the number of all multi-session and core sources;
\item achieve a near-uniform distribution of all and in particular core sources over the sky;
\item substantially improve the source position uncertainties and accuracy;
\item achieve a near-uniform distribution of source position errors over the sky.
\end{itemize}

It is expected that these goals will be achieved in the framework of regular VLBI2010 operations.
Active participation of the Southern stations is a key point in this plan.

Finally, an ICRF4 catalog with near-uniform distribution of both sources and position errors can be used for the GAIA catalog orientation.

\vspace*{0.7cm}

\noindent {\large 5. SUMMARY}

\smallskip

The following proposals can be made towards actively improving the VLBI-based ICRF and its consistency with ITRF:
Up to now, a decision on a list selection of defining sources was made, in fact, during  ICRF computation.
A good practice would be the creation of a preliminary list of the next ICRF core (defining) sources simultaneously with a ICRF realization.
Active observation of the prospective core sources for ICRF3 should be started as soon as possible.
More such sources should be included in the regular IVS multi-baseline observing sessions like R1, R4, and T2.
Special CRF-dedicated sessions should be observed by a global network with good latitudinal and longitudinal coverage.
Secondly, an agreement on a standard set of VTRF core stations should be reached, which can be used in VLBI global solutions to tie to the ITRF.
Finally, a method for the uniform description of non-linear station movement should be developed, which allows a better description, reproduction and prediction of the actual motion.

\vspace*{0.7cm}

\noindent {\large 6. REFERENCES}

{

\leftskip=5mm
\parindent=-5mm

\smallskip

Fey, A.L., Ma, C., Arias, E.F., et al., 2004, \aj, 127(6), 3587--3608.

Ma, C., Arias, E.F., Eubanks, T.M., et al., 1998, \aj, 116(1), 516--546.

Ma, C., Arias, E., Bianko, G., et al., 2009, IERS Technical Note 35, A.Fey, D.Gordon, C.S.Jacobs (eds.).

MacMillan, D., Cohen, S., 2004, In: IVS 2004 General Meeting Proc., N.R.Vandenberg, K.D.Baver (eds), 491--495.

Nikolaidis, R., 2002, Ph.D. Thesis, University of California, San Diego.

Petrov, L, 2005, In: Proc. 17th Working Meeting of EVGA, M.Vennebusch, A.Nothnagel (eds), 113-117.

Sokolova, Ju., Malkin Z., 2007, \aa, 474(2), 665--670.

}

\end{document}